\documentclass{aa}
\usepackage{natbib}
\usepackage{graphicx}
\usepackage{amssymb}

\bibpunct{(}{)}{;}{a}{}{,}

\begin{document}

\title{Angular momentum evolution for galaxies in a $\Lambda$-CDM scenario}

\author{Susana E. Pedrosa\inst{1} 
        \and Patricia B. Tissera\inst{1,2,3}}

\offprints{S. E. Pedrosa}

\institute{ Instituto de Astronom\'{\i}a y F\'{\i}sica del Espacio, CONICET-UBA, Casilla de Correos 67, Suc. 28,  C1428ZAA, Ciudad Aut\'onoma de Buenos Aires, Argentina.
\email{supe@iafe.uba.ar}
\and Departamento de Ciencias Fisicas, Universidad Andres Bello,
Av. Republica 220, Santiago, Chile.
\and Millennium Institute of Astrophysics, Av. Republica 220, Santiago, Chile.\\
}

\date{Received / Accepted}

\abstract
{Galaxy formation  in the current cosmological paradigm is a very complex process in which inflows, outflows, interactions and mergers are common events. These processes can redistribute the
angular momentum content of baryons. Recent observational results suggest that disc formed conserving angular momentum while elliptical galaxies, albeit losing angular momentum, determine a correlation between the specific angular momentum of the galaxy and the stellar mass. These observations provide stringent constraints for galaxy formation models in a hierarchical clustering scenario.}
{We aim to analyse the specific angular momentum content of the disc and bulge components as a function of virial mass, stellar mass and redshift. We also estimate the size of the simulated galaxies and confront them with observations.}
{ We use cosmological hydrodynamical simulations  that include an effective, physically-motivated Supernova feedback which is able to regulate the star formation in haloes of different masses.
We analyse the morphology and formation history of a sample of galaxies in a cosmological simulation by performing a bulge-disc decomposition of the analysed systems and their progenitors. We estimate the angular momentum content of the stellar and gaseous discs, stellar bulges and total baryons. }
{In agreement with recent observational findings, our simulated galaxies have   disc and spheroid components whose specific angular momentum contents determine  correlations with the stellar and dark matter masses with the same slope, although  the spheroidal components are off-set by a fixed fraction. The average angular momentum efficiency for the  simulated discs is $\eta \sim1$ while bulges is $\eta \sim 0.10-0.20$. For the simulated sample, the correlations found for the specific angular momentum content as a function of virial mass or stellar mass are found not to evolve significantly with redshift (up to $z\sim2$).
Both dynamical components seem to move along the correlations as they evolve. The total specific angular momentum of galaxies occupy different
positions filling the gap between pure rotational-dominated and dispersion-dominated systems. The scaling relations derived from the
simulated galaxies determine a similar relation with the virial radius, in agreement with recent observations.
}


\titlerunning{Angular momentum evolution for galaxies}
\authorrunning{Pedrosa et al.}

\maketitle

\section{Introduction}

The understanding of the morphology of galaxies and its relation with their physical properties has evolved during the last decades as new insights are provided by observations. It is well-known that the angular momentum content correlates with the morphology of galaxies along the Hubble sequence.
\citet{fall1983} has found evidences suggesting that the specific angular momentum content of both disc and elliptical galaxies holds a clear correlation with the stellar mass, but with the elliptical systems it shifts to lower values. Recently \citet{romanowsky2012}  have analysed a larger database of galaxies with different morphologies, revisiting and confirming \citet{fall1983}'s findings. These authors estimated the specific angular momentum of the stellar component $j_{\rm s}$ using kinetic and photometric data and the stellar mass obtained from near infrared luminosities. Using this larger database, they found that not only disc-dominated galaxies follow this relation but also elliptical ones. However, the latter has lower $j_{\rm s}$ at a given $M_{\rm star}$. The authors suggested that observed galaxies of all morphological types lie along almost parallel sequences with exponents $\alpha \sim 0.6$ in the $j_{\rm s}-M_{\rm star}$ diagram, but with an off-set between late and early type.   More recently, \
citet{Fall2013} showed that galaxies position in the $j_{\rm s}-M_{\rm star}$ diagram is correlated with their position in the disc-to-bulge-ratio-$M_{star}$ plane. They concluded that the description of galaxies in terms of $j_{\rm s}-M_{\rm star}$ may be regarded as a physically motivated alternative to the traditional morphological classification. The detected  $j_{\rm s}-M_{\rm star}$ relation for spheroidal galaxies suggests an angular momentum efficiency of $10-20\%$ regardless of $M_{\rm star}$  \citep{Fall2013}.

In a hierarchical clustering universe, protogalaxies acquire their angular momenta via tidal torques due to small density fluctuations \citep{peebles1969}. As baryons are initially drawn by the same tidal torques as the dark matter haloes, they are expected to acquire similar specific angular momenta.
The specific angular momentum of the dark matter haloes is supposed to grow as $M_{\rm vir}^{2/3}$. Hence, if baryons collapse conserving their angular momentum, they should also determine a similar relation. 
 Based on this hypothesis, \citet{fall1980} postulated the so-called standard  disc model (SDM). This model was able to explain the observed correlation between the angular momentum  of discs and the stellar masses \citep{MMW98}.

Several numerical works have shown that if a disc component is able to assemble then it does so conserving its specific angular momentum. Although within the hierarchical scenario for galaxy formation, this might imply a combination of processes which could contribute to retain or balance the angular momentum content so that discs are formed as expected from the SDM, \citep[e.g.][]{Agertz2011,brook2011,deRossi2012,Padilla2014,Nelson2015}.  In particular, SN feedback plays a critical role in determining the fraction of disc components, as well as mergers and interactions \citep[e.g][]{dominguez1998,Scan2009,guedes2011,Sales2012,Pedrosa2014,Lagos2015}.
Elliptical galaxies  are expected to be affected by major mergers or repetitive minor ones. This tends to decrease the angular momentum  content of the material which ends up dominated by velocity dispersion\citep[e.g.][]{meza2003,parry2009,hopkins2010,bournaud2011,sales2010,Pedrosa2014}. Internal disc instabilities could also contribute to  shape bulges and elliptical galaxies \citep[e.g.][]{efstathiou1982}.

As the final morphology of a galaxy is directly linked to its angular momentum content, there must also be  a connection with its characteristic size: when angular momentum is conserved, more extended structures are able to form. Recently, \citet{kravtsov2013} has found that characteristic sizes of stellar and gas distributions in galaxies scale approximately linearly with the virial radius derived from the abundance matching ansatz. This author found that the relation is in agreement with expectations from  the model of \citet{MMW98}.  Interestingly, he also found that this relation is present not only for late-type galaxies but also for early-type systems. From these findings, he concluded that the angular momentum content plays a crucial role in determining the sizes of galaxies of all morphological types, not only of those dominated by rotation.

In a recent paper, we have made a thorough study of the relation between the morphology of galaxies and their formation history, \citet{Pedrosa2014}. The key role played by the angular momentum content of galaxies was studied. We found that when a disc component is able to form it does so following the expected relation from the standard disc formation model in agreement with previous works \citep[see also][]{governato2010,Sales2012}. More interestingly, in this simulation,  spheroidal components are shown to follow a similar correlation but  off-set with respect to that of  rotational supported systems. It became clear from this work that SN feedback played a major role.

In this paper, we  analyse the angular momentum content of the disc and bulge components of  the  best resolved galaxies in a $\Lambda$-CDM scenario.  \citet{Pedrosa2014} analysed a sister simulation (S230A) from the Fenix Project with a stronger feedback parameter. We explore
a run (S230D) with a slightly weaker SN feedback and a different chemical distribution of the synthetized metals, which better reproduces observational results of the metallicity gradients (Tissera et al. 2015, submitted).
This is a very important point which helps to set constraints on the strength of the SN feedback \citep[e.g.][]{gibson2013}. The simulated galaxies have also larger $D/T$ stellar ratios compared to those measured in S230A. 
 We also follow them back in time to 
establish the possible existence of a redshift dependence.

This work is organized as follows: In Section \ref{sec:simus}, we describe the initial conditions and free parameters used to run the analysed simulation and  the methods used to identify and decompose galaxies. In Section \ref{sec:angmom}, we analyse the angular momentum content of our galaxies and their dynamical components. In Section \ref{sec:size}, we study  the characteristic sizes of the galaxies, and the connection with their angular momentum content. Finally, in Section \ref{sec:conclu}, we summarize our main findings.

\section{Numerical Experiments}
\label{sec:simus}

We used a version of the  code {\small GADGET-3}, an update of {\small GADGET-2 } \citep{springel2003, springel2005}, optimized for massive parallel simulations of highly inhomogeneus systems. This version of {\small GADGET-3} includes treatments for metal-dependent radiative cooling, stochastic star formation (SF), chemical enrichment, and the multiphase model for the interstellar medium (ISM) and the Supernova (SN) feedback scheme of \citet{scan2005,scan2006}. 
The SN feedback model is able to successfully trigger galactic mass-loaded winds without introducing mass-scale parameters.

Our code considers energy feedback by Type II (SNII) and Type Ia (SNIa) Supernovae. These processes  are grafted into a multiphase model for the ISM, both developed by \citet{scan2006}.
  The energy released by each SN event   is distributed between the cold and the hot phases of the  simulated ISM. The adopted multiphase model  allows the coexistence of diffuse and dense gas phases. In this model, each gas particle defines the cold and hot phases by using a local entropy criteria. This definition allows particles to decouple hydrodynamically from particular low-entropy ones if they are not part of a shock front. Each cold gas particle has a reservoir where injected SN energy is stored until they fullfil the conditions to join their local hot phase. It is  worth mentioning that this SN feedback scheme does not include parameters that  depend on the global properties of the given galaxy (e.g. the total mass, size), thus making it suitable for cosmological simulations where systems with different masses have formed in a complex way. Our code also includes the chemical evolution model developed by \citet{mosco2001}. SNII and SNIa contribute with chemical elements estimated by adopting the 
chemical yields of \citet{WW95} and \citet{iwamoto1999}, respectively. 

As mentioned before, for this study, we used one of the Fenix Project's simulations (S230D). These share the same initial conditions but assume different parameters for the SF and SN feedback sub-grid physics.
The initial conditions are consistent with the concordance model with $\Omega_{\Lambda}=0.7$, $\Omega_{\rm m}=0.3$, $\Omega_{b}=0.04$, a normalization of the power spectrum of $\sigma_{8}=0.9$ and $H_{0}= 100 h \ {\rm km} \ {\rm s}^{-1}\ {\rm Mpc}^{-1}$, with $h=0.7$. 
The simulated volume represents a box of $10$ Mpc h$^{-1}$ comovil with $2 \times 230^3$ initial particles, achieving a mass resolution of $5.9\times 10^{6}{\rm h^{-1}\ M_{\odot}}$ and $9.1\times 10^{5}{\rm  h^{-1}\  M_{\odot}}$ for the dark matter and initial gas particles, respectively. The gravitational softening is $0.5$~kpc~h$^{-1}$.   We acknowledge the fact that the initial condition represents a small volume of the Universe. It  was chosen to represent a typical region of the Universe with no large structure nearby. Using the S230A run, \citet{deRossi2013}  verified that haloes' growth is well-reproduced by confronting the results with those obtained from the Millennium simulation by \citet{fakhouri2010}. The S230A was also analysed by  \citet{Pedrosa2014} to study the morphology and angular momentum content of the simulated galaxies. Recently, \citet{artale2015} introduced the effects of high-mass X-ray binary feedback in the very early Universe in
a counterpart run of S230D.

Simulation S230D used a higher gas density threshold for star formation and a  lower energy per SN event ($0.7 \times 10^{51 }$~erg) than in S230A \citep[e.g.][]{deRossi2013,Pedrosa2014}. Simulation 230D injects a larger fraction of new synthetized chemical elements
 into the cold phase, while the fraction of SN energy is equally distributed between the cold and hot phases (S230A injected equal fraction
of energy and chemical elements into the cold and hot phases close to the SF region). This different combination 
of SF, SN feedback parameters  and metal distribution produced a good representation of chemical abundances on the disc components (Tissera et al., submitted).

\subsection{The simulated galaxy sample}
\label{sec:sample}

\begin{figure*}
\centering
\resizebox{5cm}{!}{\includegraphics{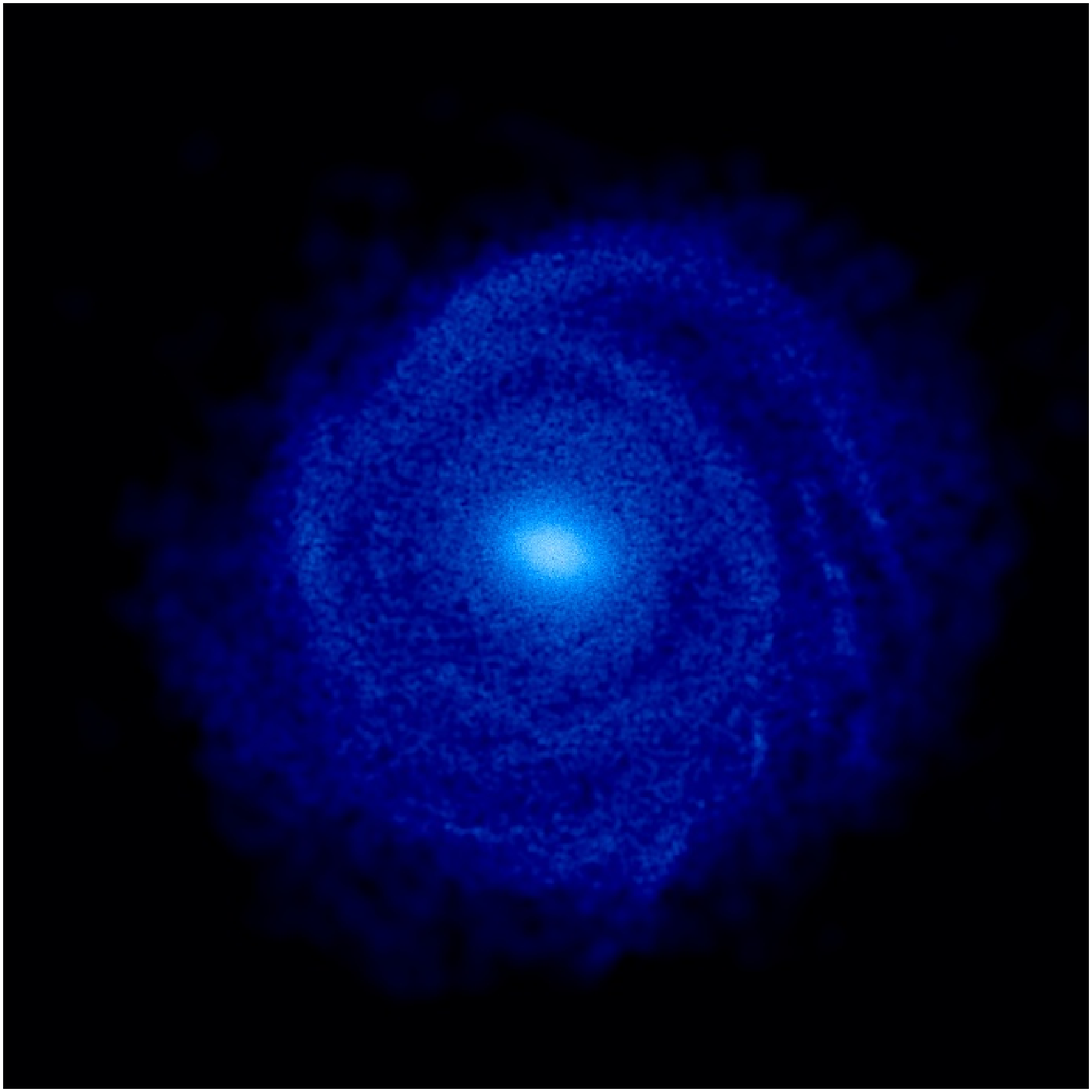}}
\hspace*{-0.2cm}
\resizebox{5cm}{!}{\includegraphics{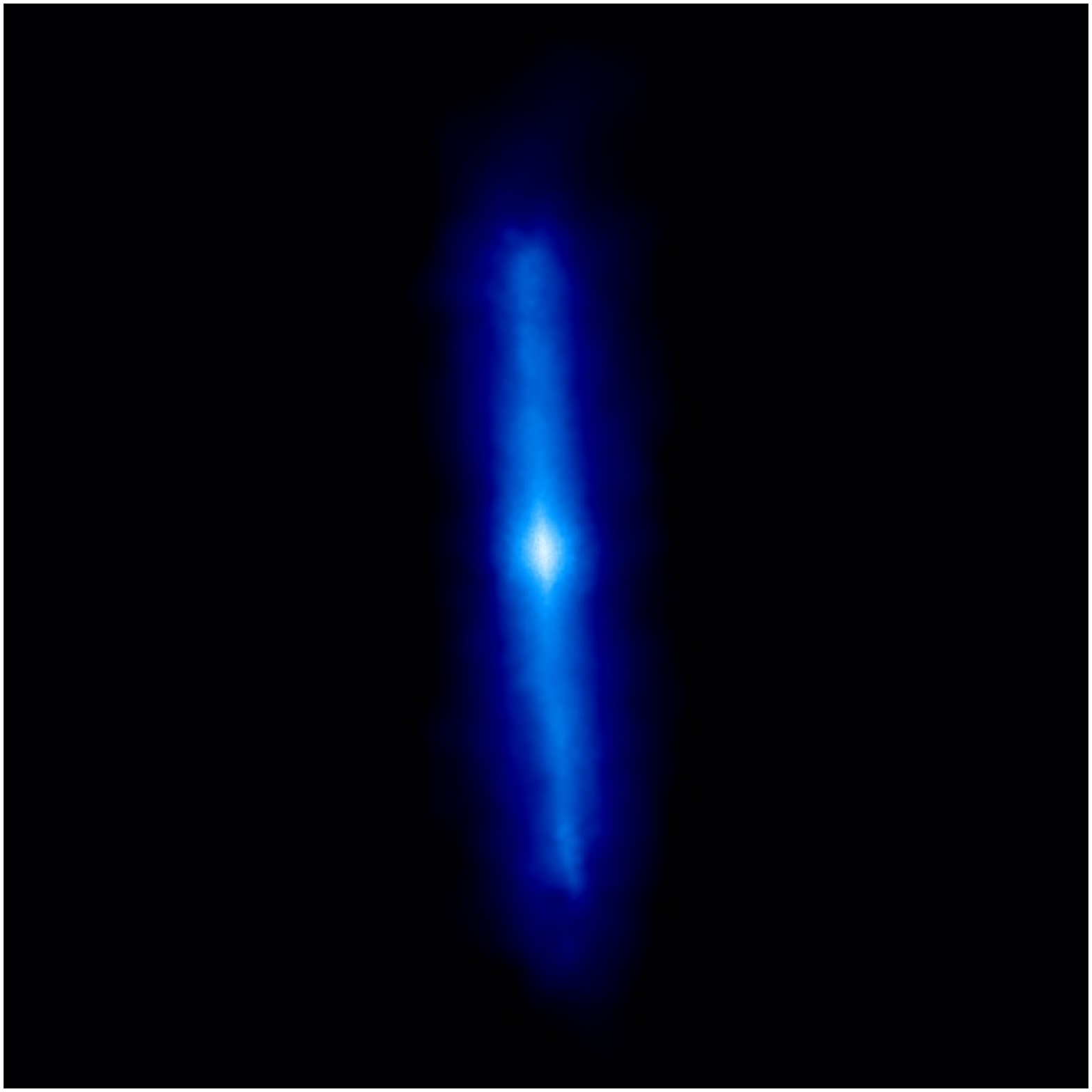}}\\
\hspace*{-0.2cm}
\resizebox{5cm}{!}{\includegraphics{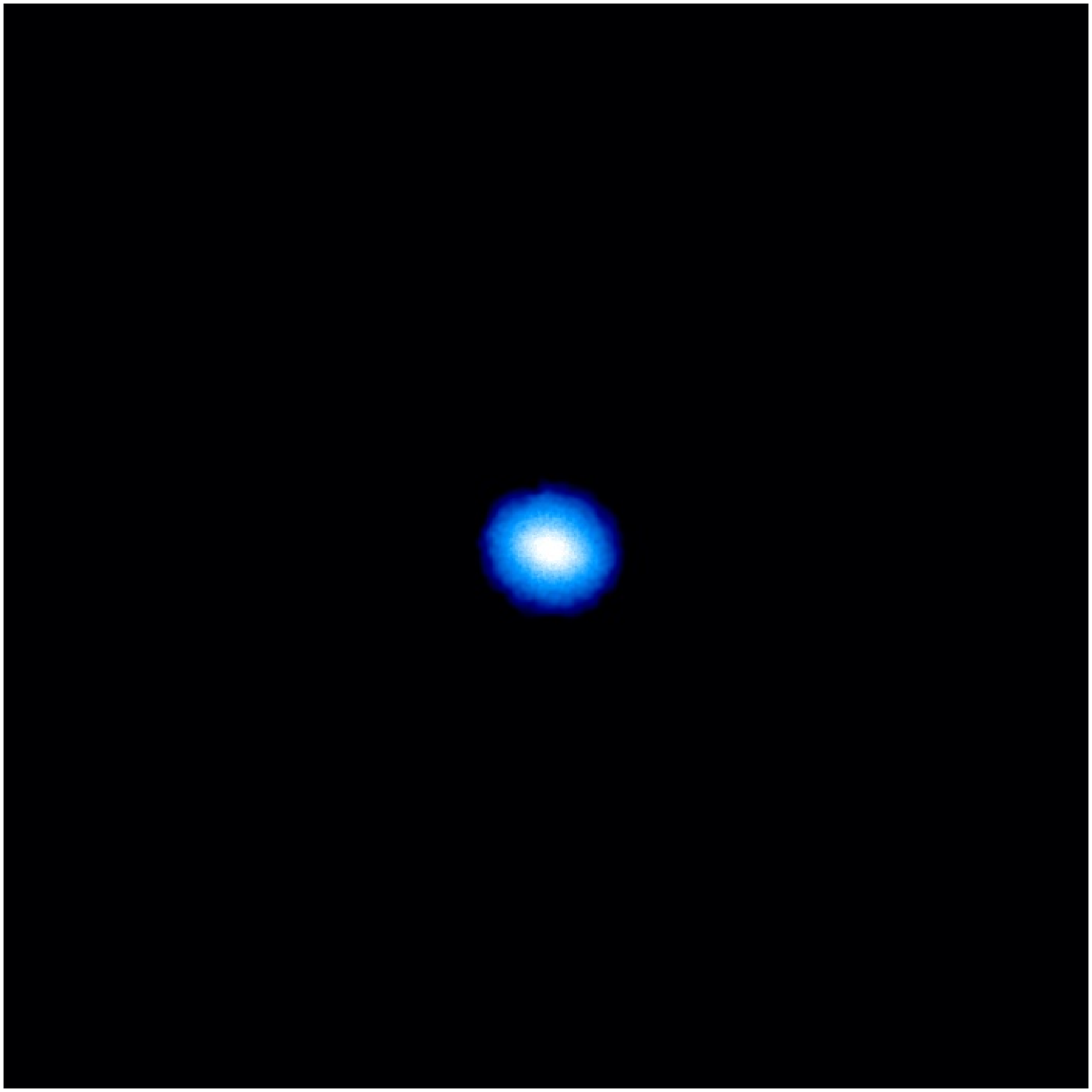}}
\hspace*{-0.2cm}
\resizebox{5cm}{!}{\includegraphics{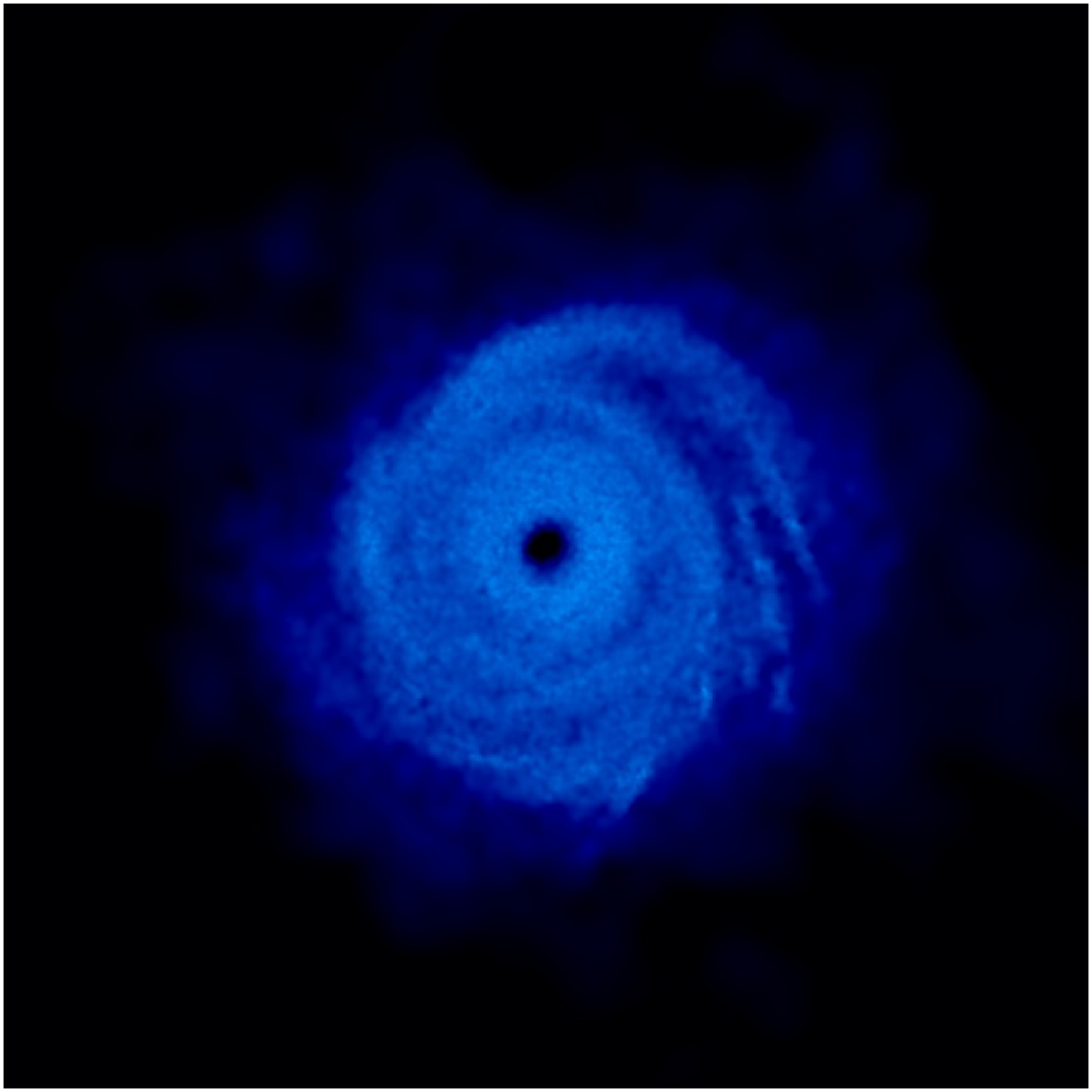}}
\hspace*{-0.2cm}
\caption{Upper panels: Density map for a face-on and an edge-on view of a late type galaxy from the 230D simulation catalog. Lower panels: Disc and central spheroid components.}
\label{fig:Map_dens}
\end{figure*}

Our galaxy sample was built up   by using a Friends-of-Friends algorithm to identify the virialized structures. Then, the substructures were discerned by using the SUBFIND scheme \citep{springel2001}, which iteratively determines the self-bound substructures within the virial radius, $r_{200}$. {\bf For illustration purposes, an example of a late type galaxy of our simulated catalog is shown in Fig.~\ref{fig:Map_dens}, (upper panels). }

To diminish resolution issues, we restricted our study to systems with a total number of particles greater than $2000$ in  the main  galaxy (which corresponds to $M_{\rm vir} \sim 10^{10.3} M_{\odot}$).  We also tested that the estimated relations are conserved
when this particle number limit is increased to $\sim 10000$.

In order to classify our simulated galaxies into early-type and late-type systems,  we estimated the parameter $\epsilon$ of the  star (gas) particles defined as $\epsilon = J_{\rm z} /J_{\rm z,max}(E)$, where $J_{\rm z}$ is the angular momentum component in the direction of the total angular momentum, and $J_{\rm z,max}(E)$ is the maximum $J_{\rm z}$ over all particles at a given binding energy, $E$. A particle on a perfect prograde circular orbit in the disc plane should have  $\epsilon=1$. 
We considered particles with $\epsilon$ higher than $0.5$ to be part of a disc. The particles  that did not satisfy this requirement were considered to belong to the spheroidal component. By inspecting the $\epsilon - E$ plane, we checked whether the adopted limits were suitable to individualize the rotationally supported components.  This threshold is less restrictive than those usually adopted 
\citep[e.g.][]{tissera2012}, so if there were stars in a thick disc, they would contribute by decreasing the signals for angular momentum conservation, for example.
 We also distinguished between the central spheroid and the stellar  halo components depending on their bounding energy following \citet{tissera2012}. In Fig.~\ref{fig:Map_dens} (lower panels)  we display the disk and central spheroid components for the galaxy previously shown, after the descomposition.  

Therefore, for each simulated galaxy, we determine the  disc (i.e.rotational-dominated) and the central spheroids components (i.e.dispersion-dominated) based on these dynamical criteria. This procedure  was also applied to galaxies identified at $z = 1$ and $z = 2$ in order to study the evolution of the angular momentum content in the galaxies,
their central spheroids and disc components and the dark matter haloes.

\section{Angular Momentum }
\label{sec:angmom}

\begin{figure*}
\centering
\resizebox{7cm}{!}{\includegraphics{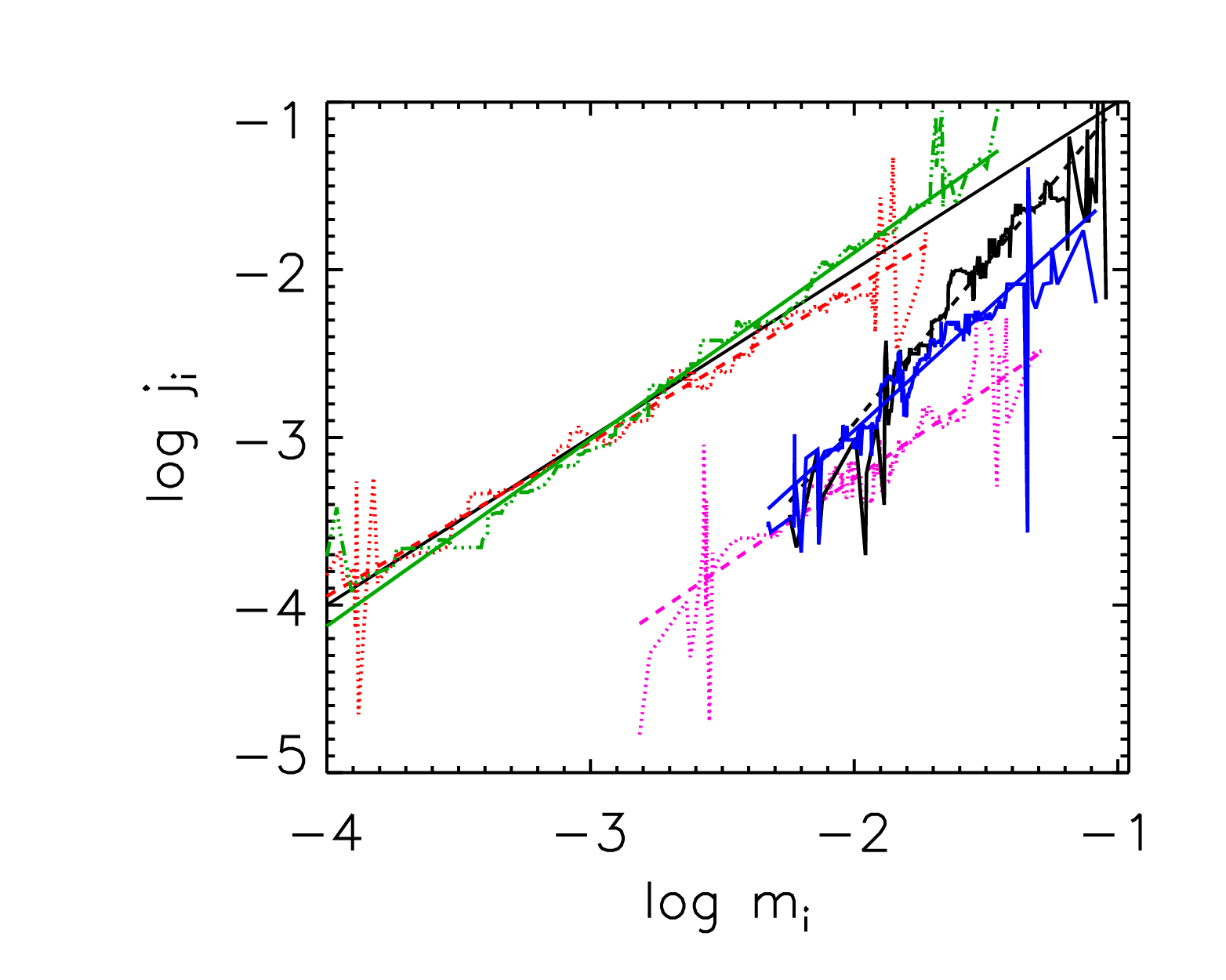}}
\resizebox{7cm}{!}{\includegraphics{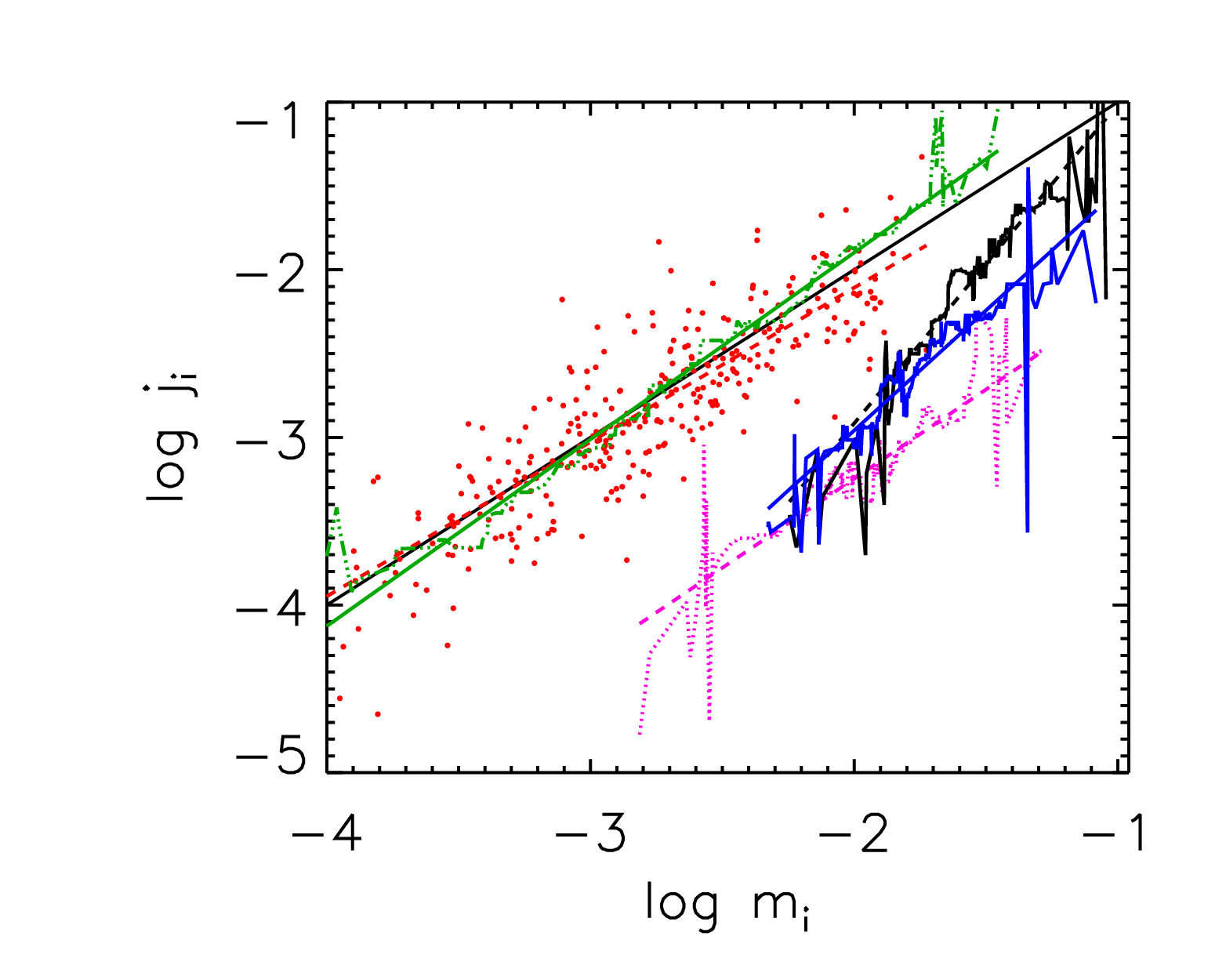}}\\
\resizebox{7cm}{!}{\includegraphics{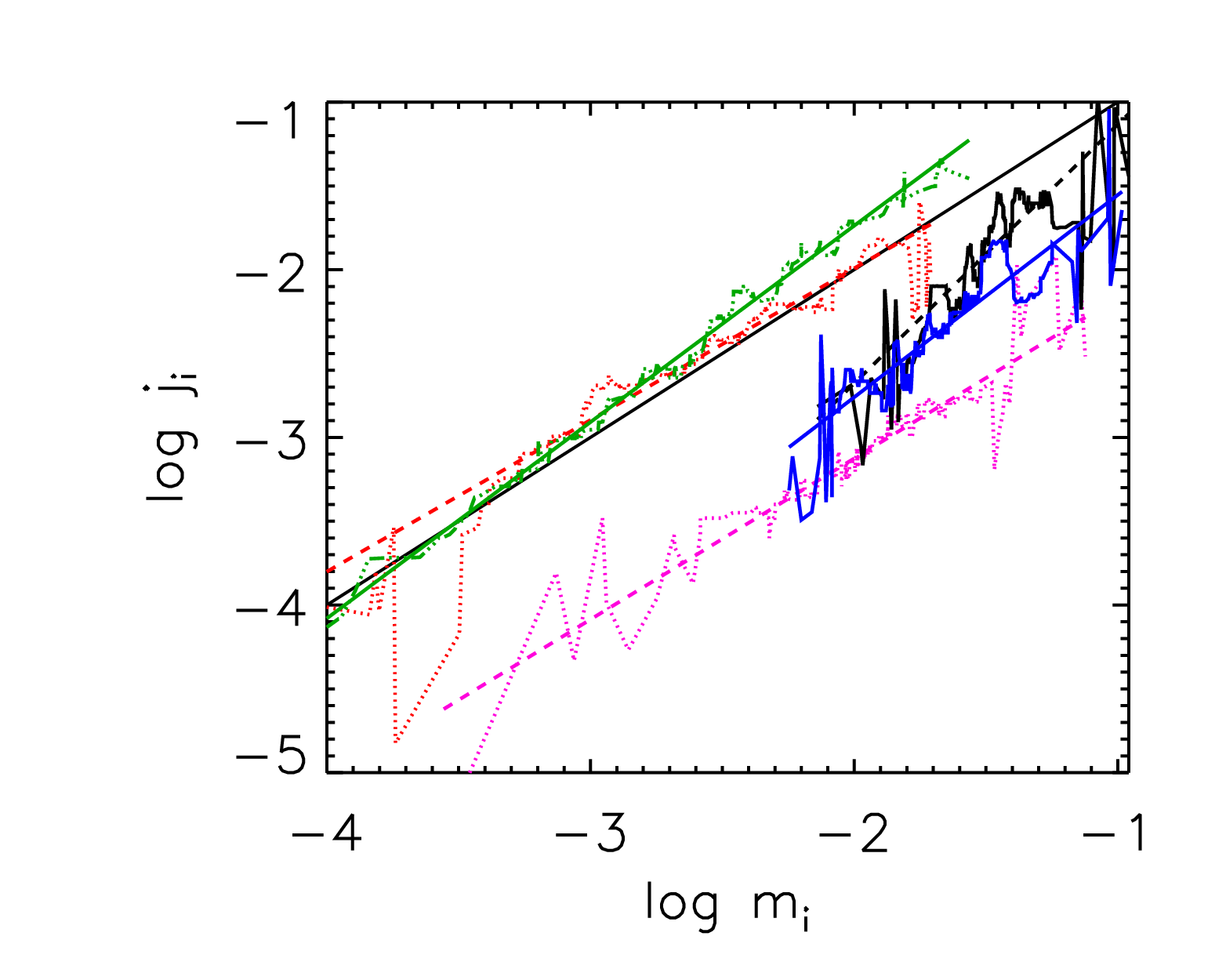}}
\resizebox{7cm}{!}{\includegraphics{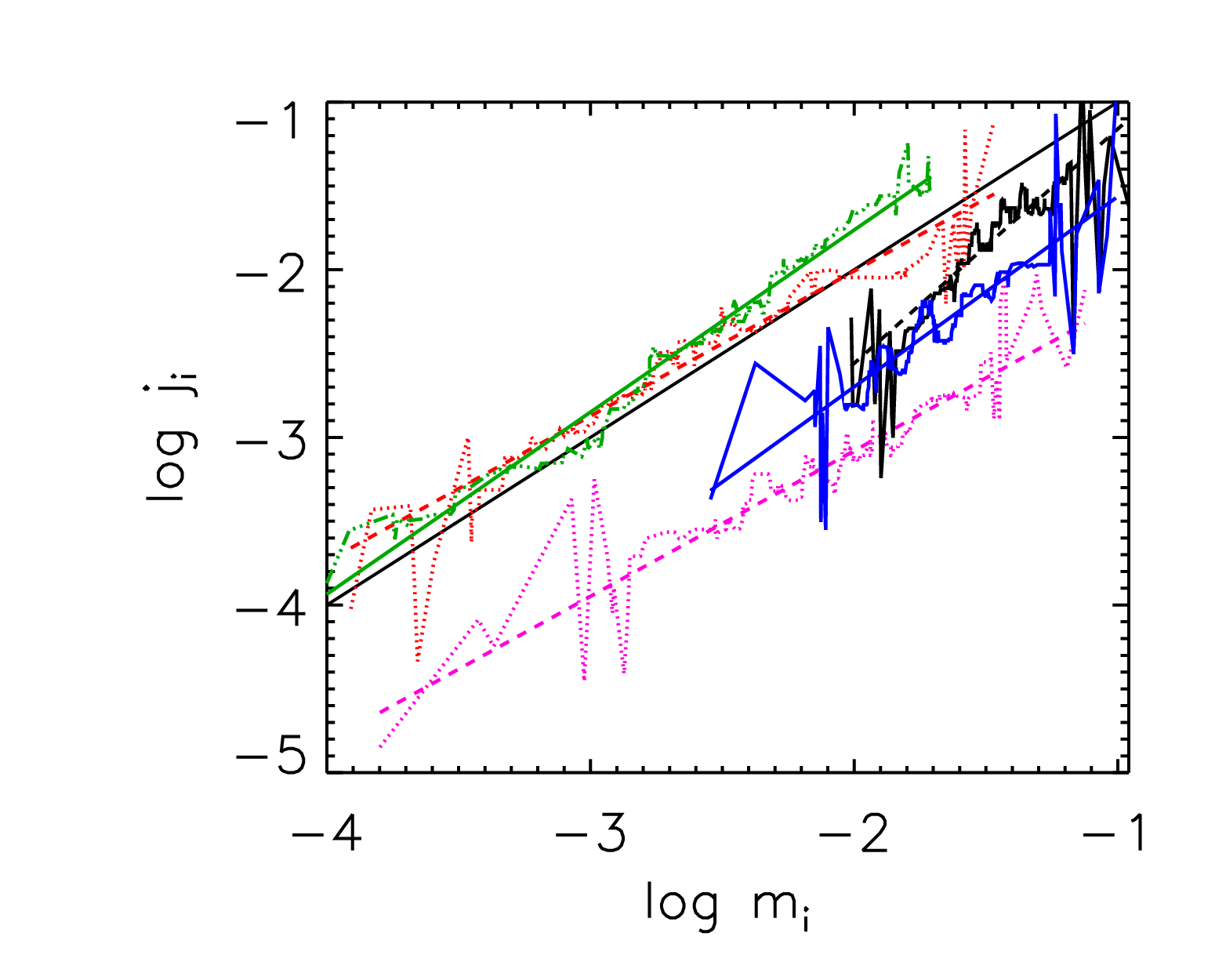}}
\hspace*{-0.2cm}
\caption{Relation $j_{\rm i}-m_{\rm i}$ medians for the stellar (red dashed lines) and gaseous (green dashed lines) disc components, the stellar spheroidal components (pink dashed lines), the total stellar component (blue line) and the total baryonic component (black lines). The solid black line represents the relation in the case of angular momentum conservation. The stellar and gas disc components are consistent with angular momentum conservation while the spheroidal components  determine a relation with a similar slope but off-set by a constant factor suggesting that spheroids have lost a similar fraction of angular momentum with respect to the total angular momentum of their virial haloes. The relation has been estimated for $z=0$ (upper panel, in the right one we include the scatter data for the stellar disc component), and for $z=1$ and $z=2$ (bottom panel, left and right respectively), finding similar trends. }
\label{fig:jdmd_D}
\end{figure*}

To calculate the properties of the simulated galaxies, we adopted  the half-mass stellar ratio ($r_{\rm HM}$), defined as the radius that enclosed $50\%$ of the total stellar mass. This characteristic radius was chosen in order to make an adequate comparison with observational results. The half-mass radius is usually adopted to estimate galaxy properties in observations. In the case of \citet{kravtsov2013} for example, he uses the stellar half mass radius to estimate his $R_{hm} - R_{vir}$ relation. We estimated the ratio between  the  angular momentum of the stars (gas) in the discs ($J_{\rm D}$) within $r_{\rm HM}$  and the corresponding of the dark matter haloes within the virial radius ($J_{\rm H} $), $j_{\rm d} = J_{\rm D} /J_{\rm H} $, as a function of the ratio between the corresponding masses,  $m_{\rm d} = M_{\rm D} /M_{\rm H} $\footnote{Note that capital letters will be used for the total angular momentum and total mass and to denote the
dynamical components: disc (D), central spheroid (S) or dark matter halo (H).}. We followed the classical notation of \citet{MMW98} to define $j_{\rm d}$ and $m_{\rm d}$. Similarly, we defined $j_{\rm s}$ and $m_{\rm s}$ ratios by using the star particles identified to belong to the central spheroidal components. Those corresponding to the whole stellar mass and the total baryonic mass were also estimated and analysed. These relations can be seen  in Fig.~\ref{fig:jdmd_D} for $z=0$ (upper panel), $z=1$ (middle panel) and $z=2$ (lower panel).  As shown, there is a clear correlation for the disc and central spheroidal components  in the three redshifts \citep[see also][]{dominguez1998,Pedrosa2014}.

Let us first analyse the relations at $z=0$.  As can be seen in Fig.~\ref{fig:jdmd_D}, there is a one-to-one relation between $j_{\rm i}$ and $m_{\rm i}$, where $\rm i$ represents the gas or stellar discs components as expected in the case of angular momentum conservation.   As discussed before, the one-to-one correlation is the consequence of the fact that all the material in the system experienced the same external torques before it collapsed to form a bounded system, and that the collapse occurred with angular momentum conservation.
 Interestingly, for the stellar central spheroidal components a clear correlation is also detected but off-set from that
determined by the disc components. This indicates that
the material which ended up dominated by dispersion lost angular momentum as expected,  but it did so by a similar fraction, regardless of the  final stellar fraction, $m_{\rm i}$. This trend was previously shown in \citet[][ figure 3]{Pedrosa2014} and in \citet[][see also \citet{dominguez1998} for earlier results with lower numerical resolution and a different numerical code and subgrid physics]{Pedrosa2015}. 

In Fig.~\ref{fig:jdmd_D} the same relations are estimated at $z=1$ and $z=2$. As can be seen, both the stellar and gaseous discs are always consistent with the one-to-one relation. However, the most interesting result is again provided by the spheroidal components which determine a similar correlation as a function of redshift. Regardless of their formation history, the disc and spheroidal components 
 determine   relations with the stellar and dark matter masses which are  statistically unchanged up to $z \sim 2$.  The gap between the angular momentum content of the galaxy respect to that of the dark matter  does not show a systematic change with redshift.

Hence, our simulated galaxies have disc and central spheroidal components consistent with the relation $ j_{\rm i} = A \times  m_{\rm i}^{\alpha} $, where $A  \sim 1$ for the disc components and
$ A \sim 0.10-0.20$ for the central spheroidal components.
We fit a linear regression to the log-log relations finding slopes of $\alpha \sim 0.9 $ and $\alpha \sim 1.1 $ for the stellar discs and central spheroids, respectively at $z=0$.

These relations are found not to evolve significantly with redshift at least up to $z \sim 2$ ($\alpha \sim 0.90,0.87$ and $\alpha \sim 0.96, 0.86$ for the stellar discs and spheroids at $z=1$ and $z=2$, respectively) 
Therefore, the spheroidal components, albeit dominated by dispersion, retain a similar fraction of the angular momentum imprinted by tidal torques onto the  small perturbations
during the first stages of the evolution of the structure \citep{peebles1969}, regardless of their assembly histories or  $m_{\rm i}$.

The slope of the relation $j_{\rm i}$-$m_{\rm i}$ changes when the total stellar mass (i.e. total stellar discs and bulges) or the  baryonic mass are considered for the estimation of these ratios  (blue and black lines, respectively), reaching values close to  $\alpha \sim 2 $ due to the combination of  angular momentum content of both disc and bulge components. This is in agreement with the results obtained by previous works where the angular momentum of baryons were studied without discerning between disc and spheroidal components \citep[e.g.][]{Sales2012}.  In this case, the gap is filled with the combination of contributions from both components and the slope of the mean relation gets steeper. We found indications of a flattening in the relation for the total baryonic mass with redshift ($\alpha \sim 1.54, 1.40 $ for $z=1$ and $z=2$, respectively ). However, a larger sample is needed to confirm the level of evolution since the dispersion in our sample increases with increasing 
redshift.

\begin{figure}
\resizebox{8cm}{!}{\includegraphics{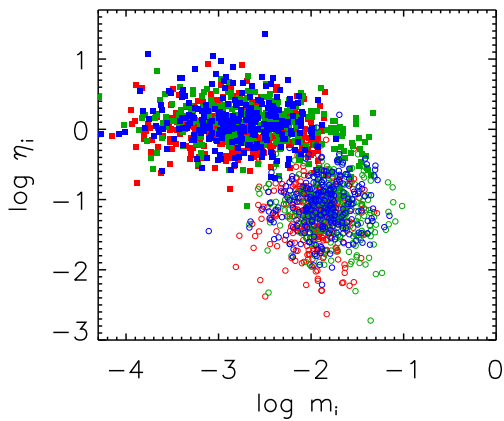}}
\hspace*{-0.2cm}
\caption{Angular momentum  efficiency  $\eta_{\rm i}$  as a function of  $m_{\rm i}$ ratio  for  disc (squares) and central spheroid (circles) components
 at $z=0$ (red), $z=1$ (green) and, $z=2$ (blue). }
\label{fig:Eta_D}
\end{figure}

We have also analysed the angular momentum  efficiency $\eta_{\rm i}$ for both components, defined as the ratio
 between  $j_{\rm D}$  ($j_{\rm S}$) of the disc component (central spheroids) and  $j_{\rm H}$ of the dark matter halo, following  \citet{sales2010}. 
In Fig.~\ref{fig:Eta_D} we  plot $\eta_{\rm i}$ versus $m_{\rm i}$ for the three
analysed redshifts. As expected, the disc components are found to have $\eta \sim 1$, on average, reflecting the conservation of angular momentum during their formation. Conversely, the  central spheroid has $\eta \sim 0.10-0.20$, on average.  
In both cases, there is no clear evolution with redshift (although the dispersion increases slightly with increasing $z$). 
The different locations of $\eta $ for the two dynamical components reflect their different evolutionary histories which  introduce dispersion but, nevertheless, take place in an organized fashion. 
The narrow range of $m_{\rm i}$ values detected for the spheroids indicate that there are not tiny bulges or massive spheroids in our simulations. Hence, our results should be verified using larger and higher-resolution simulations.

\subsection{The specific angular momenta of galaxies and their dark matter haloes}
\label{sec:angspeci}

\begin{figure}
\resizebox{8cm}{!}{\includegraphics{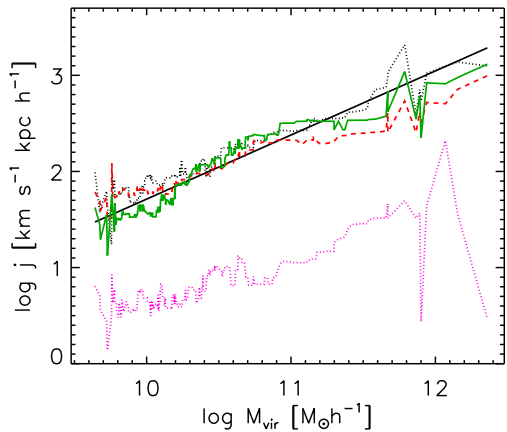}}
\hspace*{-0.2cm}
\caption{Medians of the specific stellar angular momentum of the gaseous (green line) and stellar  disc (red, dashed line) components and the central stellar spheroidal components (pink, dotted line) as a 
function of the virial mass of their host haloes. For comparison purposes, the  relation estimated  for the dark matter haloes (black dotted line)  and the expected theoretical  relation (black solid line) have been also included. }
\label{fig:jMvir_z0_D}
\end{figure}

In this section, we analyse in detail the $j$ of  the  disc and  central spheroids in relation to that of their dark matter haloes.
In Fig.~\ref{fig:jMvir_z0_D}, we show $j$   for the stellar disc and spheroidal components as a function of virial mass of their host galaxies. They define parallel relations consistent with  $ j \propto M^{2/3}$. This is in agreement with  the expected theoretical relation 
derived  under the hypothesis that the dark matter halo spin parameter is constant. 
For comparison purposes, we have also included the estimations for the gaseous disc components and the simulated dark matter haloes which
agree well with the expected theoretical relation under the hypothesis of angular momentum conservation (solid line).

Both, the spheroidal and disc components determine relations with the same slopes, which indicates that the stellar spheroids have formed from material of lower content of angular momentum (by a similar fraction), regardless of the virial mass of their host galaxy (or the relative fraction of stellar mass accumulated in the central region as shown in Fig.~\ref{fig:jdmd_D}). This trend is similar to that reported by \citet{genel2015}. This agreement is greatly encouraging considering the differences in the codes, sub-grid physics and criteria adopted to classify the simulated galaxies.

\subsection{Evolution with redshift}
\label{sec:redshift}

A question naturally arises about the possible evolution of $j$ in the disc and central spheroidal components with redshift. This might imply
a change in the slope or fraction of lost angular momentum,  or both, as  galaxies evolve.
To explore this, we estimated  the $j$ values for the stellar discs and central spheroids as a function of $M_{\rm vir}$  for $z=1$ and $z=2$. Recall that at each  redshift,  we  apply the dynamical decomposition explained in Section 2.1.

\begin{figure}
\resizebox{8cm}{!}{\includegraphics{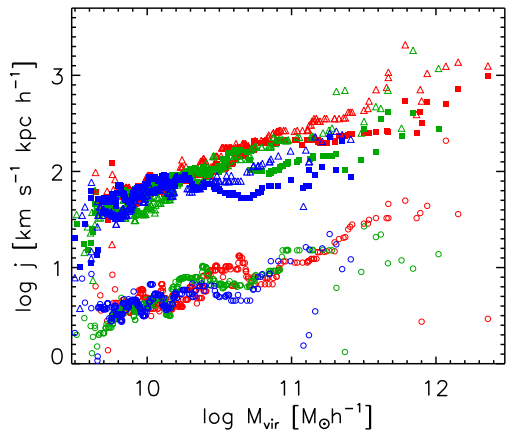}}
\hspace*{-0.2cm}
\caption{Specific stellar angular momentum for the stellar disc (empty triangles) and central spheroid (empty circles) components as a function of 
the virial mass for $z=0$ (red), $z=1$ (green) and, $z=2$ (blue). The gap between the two dynamical components does not evolve with redshift.
 For comparison purposes $j$ for the  the dark matter haloes have been also included (filled squares). Note that the
angular momenta of the discs and central spheroids are given at $r_{\rm HM}$.}
\label{fig:jBD_z_D}
\end{figure}

 As shown in Fig.~\ref{fig:jBD_z_D}, we found no significant
 evolution of these relations since $z \sim 2$. The gap between the $j$ of rotational-dominated and dispersion-dominated components remains approximately constant, implying a lost  of $80-90\%$ of the initial angular momentum content by the material which ends up in the spheroidal component, regardless of  the virial mass of the halo or of  the redshift.

This is a remarkable trend which suggests again that, independently of the assembly history of galaxies, 
their disc and bulge components will contain similar amounts of angular momentum as they evolve.

Note that in Fig. \ref{fig:jBD_z_D} we show the $j$ for the stellar disc and bulge components, independently of the global morphology of the galaxies. Hence, some of the simulated galaxies are
disc-dominated while others are spheroid-dominated systems. So, their total $j$ will be distributed filling the gap and determining a steeper mean relation as shown in
in Fig. ~\ref{fig:jdmd_D}. Therefore, these trends of conservation of the $j$ by rotational-dominated components and equally lost of $j$ by dispersion-dominated ones will be erased as the gap is filled with the combination of systems. The separation
of discs and central spheroids  is crucial
to get a deeper insight into the physical problem. Nevertheless, we can not explain the origin of the similar angular momentum
efficiency for central spheroids yet.

\subsection{Comparison with observations}
\label{sec:obser}

\begin{figure}
\resizebox{8cm}{!}{\includegraphics{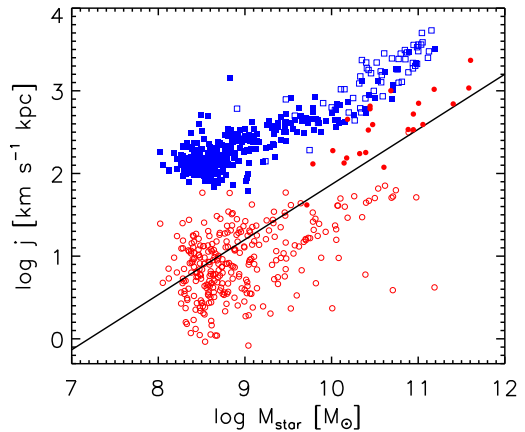}}
\hspace*{-0.2cm}
\caption{Relation between $j$  and the total stellar mass for the simulated stellar disc components (blue, filled squares) and the
stellar spheroids (red, open circles).  The observational estimations reported by \citet{Fall2013} are depicted for comparison:  disc components (blue, open squares) and Elliptical galaxies (red filled circles). The theoretical relation $j \propto M^{2/3}$ is also included for comparison (black, continuous line). }
\label{fig:jMstar_DT_D_Fall2}
\end{figure}

Recently, \citet{Fall2013} reported results which support the existence of a relation between the $j$ and stellar mass with the same slope for both ellipticals and disc dominated-galaxies, as found by \citet{fall1980}. The gap between them indicates a lost of angular momentum by the material which formed the ellipticals by a similar fraction, regardless of stellar mass. 
Our simulations seem to reproduce naturally this trend as can be deducted from Fig. ~\ref{fig:jdmd_D} and more clearly seen from 
Fig.~\ref{fig:jMstar_DT_D_Fall2} where the observational estimations from  \citet{Fall2013} are included.
As shown in this figure, the stellar disc component determine a relation which agrees well with the one estimated for observed discs.
In the case of central spheroids, they are also consistent with observations provided that stellar bulges behave
similarly to elliptical galaxies \citep[as also suggested by][]{Fall2013}. The good agreement between the 
observations and simulations suggests that the
latter are able to grasp  well enough relevant aspects of the physics at play.

In order to improve the comparison of our results with observations, we estimated the total $j$ defined by the whole stellar 
populations within the galaxy radius\footnote{The galaxy radius is defined as the one which encloses $\sim 80 \% $ of the baryonic mass.}. Previous figures show results estimated at $r_{\rm HM}$.
The morphology of a simulated galaxy is defined  by the ratio  $D/T$ between the stellar disc mass, $D$,  and the total stellar mass of the galaxy, $T=D+B$. 
Simulated galaxies are classified according to the $D/T$ in four subsamples from disc-dominated ($D/T > 0.6$) to spherodial-dominated  ($D/T < 0.2$) ones.

  We should keep in mind that the comparison should be taken as indicative since  differences between the observed and simulated parameters are always expected. For example, while observations use the gas information to trace the angular momentum content of the stellar discs under the hypothesis that they should be dynamical comparable, the simulated calculations make use of the information store directly in the stellar populations.  As we show below, the gaseous discs conserve their angular momentum and they compare well to the relation determined by Sc galaxies, where the dispersion-dominated component is small. This indicates that this type of galaxies is missing in the simulations.

\begin{figure*}
\centering
\resizebox{10cm}{!}{\includegraphics{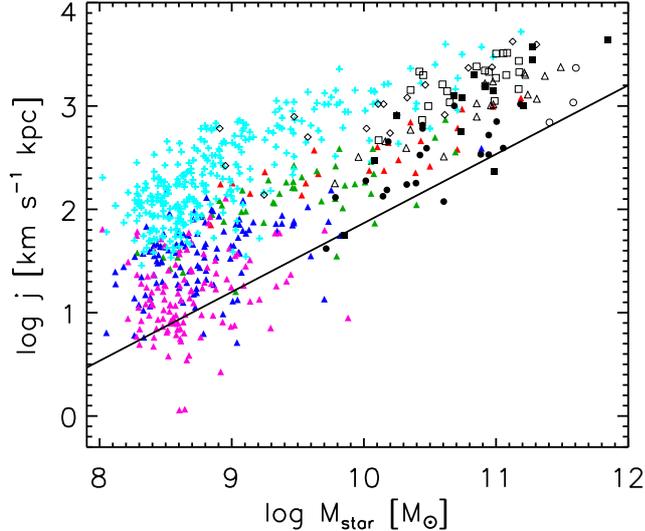}}
\hspace*{-0.2cm}
\caption{Relation between $j$  and the total stellar mass for simulated galaxies with different disc-to-bulge ratios: $D/T > 0.6$ (red, filled triangles), $0.4 < D/T < 0.6$ (green, filled triangles), $0.4 < D/T <0.2$ (blue, filled triangles) and  $D/T < 0.2$ (pink, filled triangles) calculated at the galaxy radius. The relation for the gaseous discs are also included (cyan crosses).The observational estimations reported by \citet{Fall2013} are depicted for comparison:  Sa-Sab (black triangles), Sb-Sbc (black squares), Sc-Sm (black diamonds), E (black filled circles), S0 (black filled squares, sE (black circles). The theoretical relation $j \propto M^{2/3}$ is also included (black, continuous line). }
\label{fig:jMstar_DT_D}
\end{figure*}

In Fig. \ref{fig:jMstar_DT_D}, we show $j$ as a function of galaxy stellar mass with different $ D/T$ ratios. A correlation can be seen in the sense that higher $D/T$ ratios are related to higher contents of specific angular momentum, as expected. It also presents a good agreement with the observational trend found by \citet{Fall2013}. The gap between disc-dominated systems ($D/T > 0.6$)
 and spheroidal-dominated ones ($D/T < 0.2$) is filled with the corresponding intermediate morphological types. 
Our results support the claim that this relation opens the possibility to an alternative way of mapping galaxies' morphologies. 
However, and more importantly, the agreement between the observational and numerical results suggests
that the hierarchical clustering signature is not erased in spheroidal galaxies as previously assumed. 

It is remarkable that the gap between both morphological types does not depend on the  mass of the galaxy or the virial mass of the haloes. In \citet{Pedrosa2014} a detailed study of the $D/T$ fraction was done. The evolution of the morphology was quantified by analysing the D/T ratios as a function of the redshift for  a group of 15 selected galaxies and the correlation of that parameter with their assembly history. As galaxies evolve, the $D/T$ ratios of the progenitor systems can drastically change, reflecting their different formation histories (major and minor mergers, accretions, etc). The $D/T$ fraction is a highly dynamic parameter, which indicates that the disc can be destroyed and re-built several times along its evolution, as shown in Figure 4 of \citet{Pedrosa2014}.

During galaxy assembly, angular momentum is exchanged through different physical processes. Previous works such as \citet{brook2011} have found that preferential remotion of low angular momentum material at early redshift also contributes to the redistribution of the angular momentum. They found that low angular momentum material is accreted and rapidly expelled when star formation peaks at early times, with shallow potential wells that favour early gas ejection. Later accreted material has higher angular momentum and is accreted preferentially onto the disk, contributing to re-build it. Also, gas might flow into galaxies through the dark matter filaments, feeding the galaxy with high angular momentum material, as found by \citet{dekel2009}. \citet{genel2015} studied, using a more observational approach, the angular momentum content of galaxies in the Illustris simulations. They found, in agreement with our results, that early-type galaxies retention factor is approximately 30\%. They also analised the evolution track of a group of selected galaxies, finding that different processes, such as mergers, could drastically redistribute the angular momentum content. 

Therefore, bulges are formed by important contributions of low  angular momentum material which might have lost it during the galaxy formation process \citep{Pedrosa2014}.

\section{Size relations}
\label{sec:size}

As shown in previous sections, the disc components in our simulations were formed in agreement with the hypothesis of  $j$ conservation, while the central spheroids were formed from material with lower content of $j$, by a similar fraction. From these results, we would expect a clear scaling relation between the galaxy and halo size.

The proportionality between the specific angular momentum of the baryons and that of  the halo assumed by the disc galaxy formation models would also imply  the existence of a linear relation between the galactic and virial radius of the hosting halo \citep{kravtsov2013}. The galaxy size is expected to scale as $r_{\rm gal} \propto \lambda r_{200}$ .

\citet{kravtsov2013} analysed the relation between sizes of stellar systems of galaxies characterized by their half-mass radius and the virial radius of their dark matter haloes. The latter are derived by  using the abundance matching ansatz. They found that the characteristic size of stellar and gas distributions in galaxies spanning several orders of magnitude in stellar mass, scales approximately linearly with a fraction of the the virial radius with a proportionality factor of  $\sim 0.015$. However, this author did not consider the effects of baryons on the dark matter potential well, which are naturally provided by our hydrodynamical simulations \citep[e.g.][]{Pedrosa2010, tissera2010}.   In fact, \citet{diemer2013}  claimed that due to the most common definition of the dark matter halo that considers a spherical overdensity of matter with respect to a reference density, such as the critical one or the mean matter density of the Universe, could lead to a spurious pseudo-evolution of halo mass due to 
redshift evolution of the reference density. They found that for all halo masses a significant fraction of the halo mass growth since $z \sim 1$ is due to this so-called pseudo-evolution rather than the actual physical accretion of matter. They obtained a corrected value for the size relation of $0.03 - 0.04$.
 by taking the contraction effect into account. 

\begin{figure}
\resizebox{8cm}{!}{\includegraphics{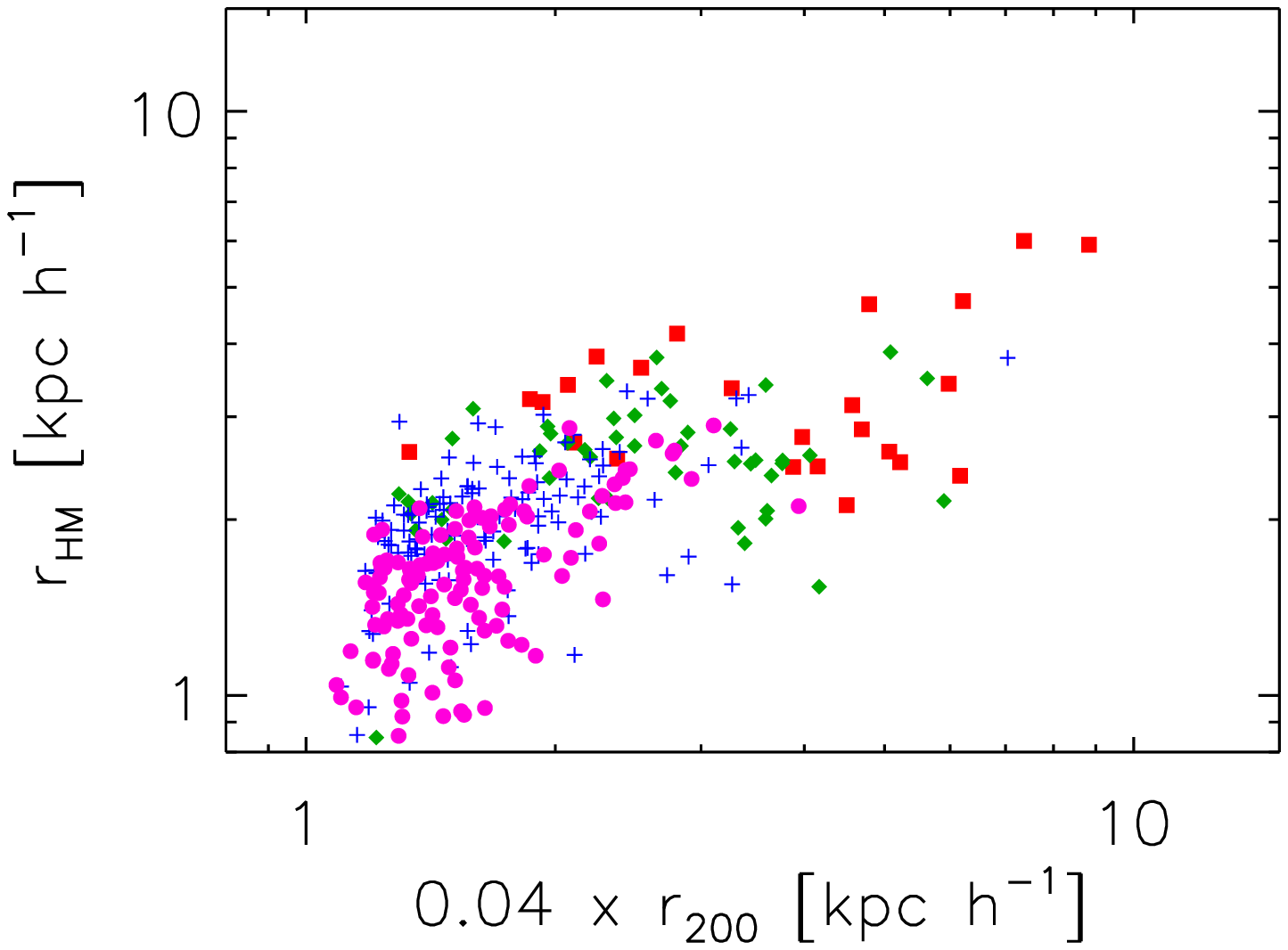}}
\resizebox{8cm}{!}{\includegraphics{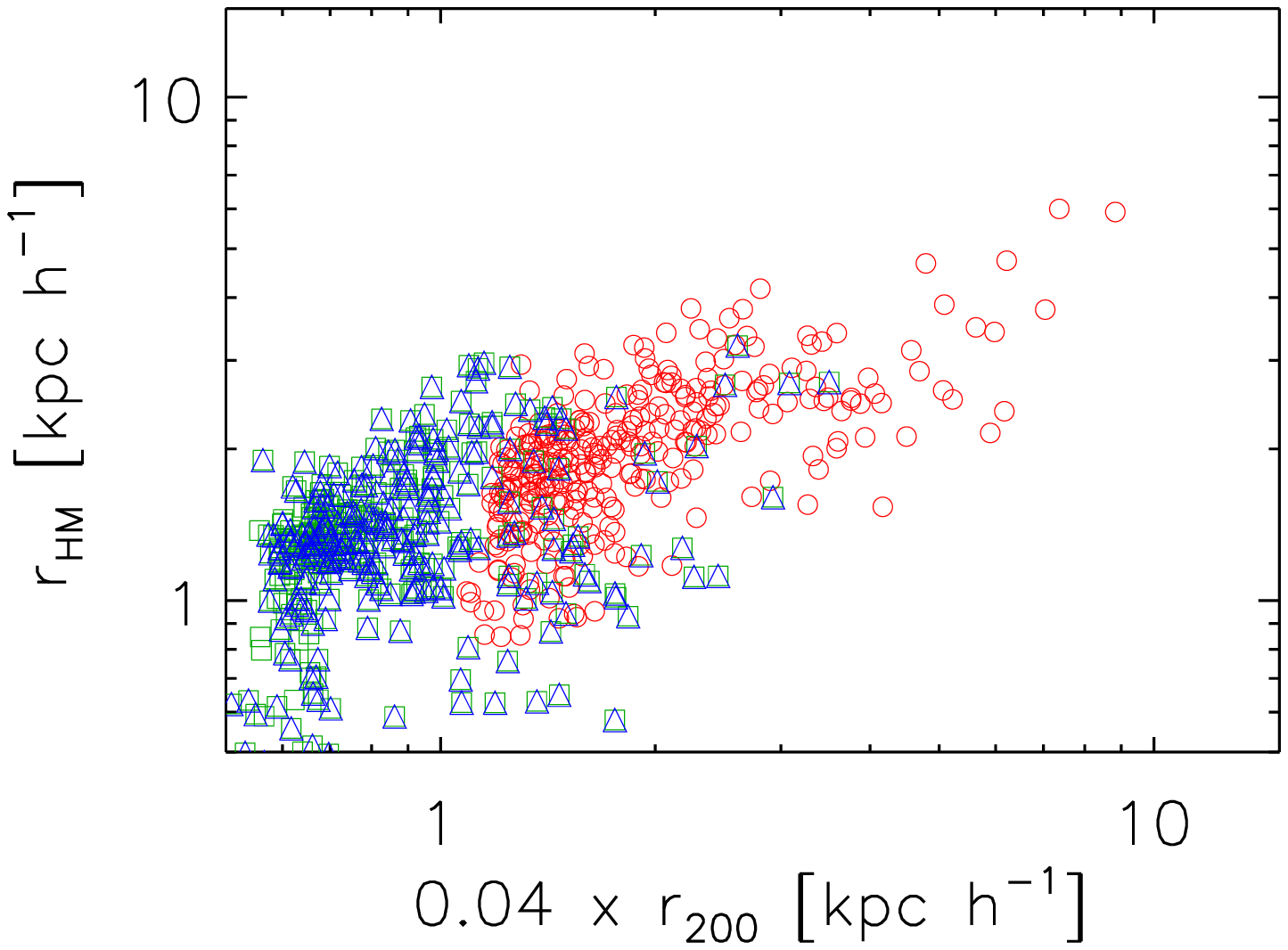}}
\hspace*{-0.2cm}
\caption{Size relations for the simulated galaxies and their host dark matter haloes: $r_{\rm HM} =0.04 \times \lambda r_{200}$. Upper panel: Simulated galaxies are colour-coded according to  the $D/T$ ratio:  $D/T > 0.6$ (red, filled squares), $0.4 < D/T < 0.6$ (green, filled diamonds), $0.4 < D/T <0.2$ (blue, crosses) and  $D/T < 0.2$ (pink, filled circles). Lower panel: The colours correspond to the different redshifts, $z = 0$ (red circles), $z = 1$ (green squares), $z = 2$ (blue triangles)}
\label{fig:Rhm_rvir_fracD}
\end{figure}

In Fig.~\ref{fig:Rhm_rvir_fracD} (upper panel) we show the stellar $r_{\rm HM}$ as a function of a fraction of $r_{200}$ for our simulated systems. Interestingly, we determined that the best factor relating both characteristic radii is $\sim 0.04$. 
In  this figure, we show the relation for galaxies with different $D/T$ ratios. It is clear that higher $D/T$ values correspond to more extended structures. And this is consistent with Fig.~\ref{fig:jMstar_DT_D} where it can be seen that also higher $D/T$ ratios correspond to higher values of  angular momentum content.
The size correlation is at place since $z \sim 2$ as expected since the angular momentum relation is also present and unchanged globally
since this time Fig.~\ref{fig:Rhm_rvir_fracD}(lower panel).

\section{Conclusions}
\label{sec:conclu}

We used hydrodynamical cosmological simulation to perform a comparative analysis of the angular momentum content in the disc and bulge components of the simulated galaxies. 
We focused our analysis on the the best resolved galaxies and used a dynamical criterion to identify the disc and bulge components at $z=0$ and for their progenitor galaxies at $z=1$ and $z=2$.

Our key findings are:

\begin{itemize}
\item{} We found  that both discs and bulges determine a relation in  the $j_{\rm i}-m_{\rm i}$  plane. Both the rotational-supported  and the dispersion dominated components follow a relation with the same slope ($j_{\rm i} \propto m_{\rm i}$). For the disc components, the correlation is the expected one-to-one relation in the case of angular momentum conservation \citep{MMW98}. Remarkably, the dispersion-dominated systems determine a parallel relation but off-set by an approximately constant factor.  Our simulated bulges  conserved $\eta \sim 10-20\% $ of their original angular momentum, regardless of the $D/T$ ratio of the  galaxy \citep[see also ][]{dominguez1998,Pedrosa2014}.  This is in agreement with observational findings \citep[][and referencies there in]{Fall2013}.

\item A similar analysis was carried out for the progenitor systems up to $z \sim  2$, finding that the central spheroids and discs determine similar relations. Hence,   regardless of the formation history of these galaxies, their disc and spheroidal components have angular momentum contents which determine relations with the  stellar and dark matter masses with similar slope up to $z \sim 2$. In the case of the  spheroidal components a similar off-set is detected showing an angular momentum efficiency of $\sim 0.10 - 0.20$. These values
vary slightly with redshift. A larger sample is required to assess robustly the level of evolution.

\item Regarding  the specific angular momentum, we found that the spheroidal and disc components determine relations with the same
slopes, $ j \propto M^{2/3}$. This indicates that the stellar spheroids have been formed by material of low angular momentum content or that have lost it by similar fraction during the assembly process, regardless of the virial mass of their host galaxy. We have analysed this relation at different redshifts, finding no evolution. Remarkably, this trend is suggesting that independently of virial masses, the disc and bulge components will conserve similar amounts of angular momentum as they evolve.

\item In good agreement with the trend found by \citet{Fall2013}, we found a clear correlation between the morphological type of a galaxy and its total specific angular momentum content: higher D/T ratios are related with higher contents of specific angular momentum, as expected. The gap between disc-dominated systems ($D/T > 0.6$) and spheroidal-dominated ones ($D/T < 0.2$) is filled with the corresponding intermediate morphological types. This supports the claim that this relation opens the possibility to an alternative way of mapping the morphology of galaxies. And this also implies that the hierarchical clustering signature is still present in spheroidal galaxies. It is remarkable that the gap between both morphological types does not depend on the mass of the galaxy or the virial mass of the haloes.

\item In agreement with \citet{kravtsov2013}, we found that the characteristic size of stellar distributions in the simulated galaxies scales approximately linearly with the virial radius in the log-log plane. The proportionality factor for our simulated systems is $\sim 0.04$ \citep{diemer2013}. This relation is found not to evolve with time: galaxies move along the relation as one moves to higher redshift.

\end{itemize}

\begin{acknowledgements}
The authors thanks Michael Fall for kindly providing the data electronically.
This work was partially supported by PICT 2011-0959 and PIP 2012-0396 (Mincyt, Argentina). PBT acknowledges partial support from  the Regular Grant UNAB 2014, Nucleo UNAB 2015 of Universidad Andres Bello and Fondecyt 1150334.
\end{acknowledgements}

\bibliographystyle{aa}
\bibliography{J_evol9}

\begin{thebibliography}{44}
\expandafter\ifx\csname natexlab\endcsname\relax\def\natexlab#1{#1}\fi

\bibitem[{{Agertz} {et~al.}(2011){Agertz}, {Teyssier}, \& {Moore}}]{Agertz2011}
{Agertz}, O., {Teyssier}, R., \& {Moore}, B. 2011, \mnras, 410, 1391

\bibitem[{{Artale} {et~al.}(2015){Artale}, {Tissera}, \&
  {Pellizza}}]{artale2015}
{Artale}, M.~C., {Tissera}, P.~B., \& {Pellizza}, L.~J. 2015, MNRAS, 448, 3071

\bibitem[{{Bournaud} {et~al.}(2011){Bournaud}, {Chapon}, {Teyssier}, {Powell},
  {Elmegreen}, {Elmegreen}, {Duc}, {Contini}, {Epinat}, \&
  {Shapiro}}]{bournaud2011}
{Bournaud}, F., {Chapon}, D., {Teyssier}, R., {et~al.} 2011, \apj, 730, 4

\bibitem[{{Brook} {et~al.}(2011){Brook}, {Governato}, {Ro{\v s}kar}, {Stinson},
  {Brooks}, {Wadsley}, {Quinn}, {Gibson}, {Snaith}, {Pilkington}, {House}, \&
  {Pontzen}}]{brook2011}
{Brook}, C.~B., {Governato}, F., {Ro{\v s}kar}, R., {et~al.} 2011, \mnras, 415,
  1051

\bibitem[{{De Rossi} {et~al.}(2013){De Rossi}, {Avila-Reese}, {Tissera},
  {Gonz{\'a}lez-Samaniego}, \& {Pedrosa}}]{deRossi2013}
{De Rossi}, M.~E., {Avila-Reese}, V., {Tissera}, P.~B.,
  {Gonz{\'a}lez-Samaniego}, A., \& {Pedrosa}, S.~E. 2013, \mnras, 435, 2736

\bibitem[{{De Rossi} {et~al.}(2012){De Rossi}, {Tissera}, \&
  {Pedrosa}}]{deRossi2012}
{De Rossi}, M.~E., {Tissera}, P.~B., \& {Pedrosa}, S.~E. 2012, \aap, 546, A52

\bibitem[{{Dekel} {et~al.}(2009){Dekel}, {Sari}, \& {Ceverino}}]{dekel2009}
{Dekel}, A., {Sari}, R., \& {Ceverino}, D. 2009, \apj, 703, 785

\bibitem[{{Diemer} {et~al.}(2013){Diemer}, {More}, \& {Kravtsov}}]{diemer2013}
{Diemer}, B., {More}, S., \& {Kravtsov}, A.~V. 2013, \apj, 766, 25

\bibitem[{{Dom{\'{\i}}nguez-Tenreiro}
  {et~al.}(1998){Dom{\'{\i}}nguez-Tenreiro}, {Tissera}, \&
  {S{\'a}iz}}]{dominguez1998}
{Dom{\'{\i}}nguez-Tenreiro}, R., {Tissera}, P.~B., \& {S{\'a}iz}, A. 1998,
  \apjl, 508, L123

\bibitem[{{Efstathiou} {et~al.}(1982){Efstathiou}, {Lake}, \&
  {Negroponte}}]{efstathiou1982}
{Efstathiou}, G., {Lake}, G., \& {Negroponte}, J. 1982, \mnras, 199, 1069

\bibitem[{{Fakhouri} {et~al.}(2010){Fakhouri}, {Ma}, \&
  {Boylan-Kolchin}}]{fakhouri2010}
{Fakhouri}, O., {Ma}, C.-P., \& {Boylan-Kolchin}, M. 2010, MNRAS, 406, 2267

\bibitem[{{Fall}(1983)}]{fall1983}
{Fall}, S.~M. 1983, in IAU Symposium, Vol. 100, Internal Kinematics and
  Dynamics of Galaxies, ed. E.~{Athanassoula}, 391--398

\bibitem[{{Fall} \& {Efstathiou}(1980)}]{fall1980}
{Fall}, S.~M. \& {Efstathiou}, G. 1980, \mnras, 193, 189

\bibitem[{{Fall} \& {Romanowsky}(2013)}]{Fall2013}
{Fall}, S.~M. \& {Romanowsky}, A.~J. 2013, \apjl, 769, L26

\bibitem[{{Genel} {et~al.}(2015){Genel}, {Fall}, {Hernquist}, {Vogelsberger},
  {Snyder}, {Rodriguez-Gomez}, {Sijacki}, \& {Springel}}]{genel2015}
{Genel}, S., {Fall}, S.~M., {Hernquist}, L., {et~al.} 2015, ArXiv e-prints

\bibitem[{{Gibson} {et~al.}(2013){Gibson}, {Pilkington}, {Brook}, {Stinson}, \&
  {Bailin}}]{gibson2013}
{Gibson}, B.~K., {Pilkington}, K., {Brook}, C.~B., {Stinson}, G.~S., \&
  {Bailin}, J. 2013, A\&A, 554, A47

\bibitem[{{Governato} {et~al.}(2010){Governato}, {Brook}, {Mayer}, {Brooks},
  {Rhee}, {Wadsley}, {Jonsson}, {Willman}, {Stinson}, {Quinn}, \&
  {Madau}}]{governato2010}
{Governato}, F., {Brook}, C., {Mayer}, L., {et~al.} 2010, \nat, 463, 203

\bibitem[{{Guedes} {et~al.}(2011){Guedes}, {Callegari}, {Madau}, \&
  {Mayer}}]{guedes2011}
{Guedes}, J., {Callegari}, S., {Madau}, P., \& {Mayer}, L. 2011, \apj, 742, 76

\bibitem[{{Hopkins} {et~al.}(2010){Hopkins}, {Bundy}, {Croton}, {Hernquist},
  {Keres}, {Khochfar}, {Stewart}, {Wetzel}, \& {Younger}}]{hopkins2010}
{Hopkins}, P.~F., {Bundy}, K., {Croton}, D., {et~al.} 2010, \apj, 715, 202

\bibitem[{{Iwamoto} {et~al.}(1999){Iwamoto}, {Brachwitz}, {Nomoto},
  {Kishimoto}, {Umeda}, {Hix}, \& {Thielemann}}]{iwamoto1999}
{Iwamoto}, K., {Brachwitz}, F., {Nomoto}, K., {et~al.} 1999, ApJS, 125, 439

\bibitem[{{Kravtsov}(2013)}]{kravtsov2013}
{Kravtsov}, A.~V. 2013, \apjl, 764, L31

\bibitem[{{Lagos} {et~al.}(2015){Lagos}, {Padilla}, {Davis}, {Lacey}, {Baugh},
  {Gonzalez-Perez}, {Zwaan}, \& {Contreras}}]{Lagos2015}
{Lagos}, C.~d.~P., {Padilla}, N.~D., {Davis}, T.~A., {et~al.} 2015, \mnras,
  448, 1271

\bibitem[{{Meza} {et~al.}(2003){Meza}, {Navarro}, {Steinmetz}, \&
  {Eke}}]{meza2003}
{Meza}, A., {Navarro}, J.~F., {Steinmetz}, M., \& {Eke}, V.~R. 2003, \apj, 590,
  619

\bibitem[{{Mo} {et~al.}(1998){Mo}, {Mao}, \& {White}}]{MMW98}
{Mo}, H.~J., {Mao}, S., \& {White}, S.~D.~M. 1998, \mnras, 295, 319

\bibitem[{{Mosconi} {et~al.}(2001){Mosconi}, {Tissera}, {Lambas}, \&
  {Cora}}]{mosco2001}
{Mosconi}, M.~B., {Tissera}, P.~B., {Lambas}, D.~G., \& {Cora}, S.~A. 2001,
  \mnras, 325, 34

\bibitem[{{Nelson} {et~al.}(2015){Nelson}, {Genel}, {Vogelsberger}, {Springel},
  {Sijacki}, {Torrey}, \& {Hernquist}}]{Nelson2015}
{Nelson}, D., {Genel}, S., {Vogelsberger}, M., {et~al.} 2015, MNRAS, 448, 59

\bibitem[{{Padilla} {et~al.}(2014){Padilla}, {Salazar-Albornoz}, {Contreras},
  {Cora}, \& {Ruiz}}]{Padilla2014}
{Padilla}, N.~D., {Salazar-Albornoz}, S., {Contreras}, S., {Cora}, S.~A., \&
  {Ruiz}, A.~N. 2014, \mnras, 443, 2801

\bibitem[{{Parry} {et~al.}(2009){Parry}, {Eke}, \& {Frenk}}]{parry2009}
{Parry}, O.~H., {Eke}, V.~R., \& {Frenk}, C.~S. 2009, \mnras, 396, 1972

\bibitem[{{Pedrosa} \& {Tissera}(2015)}]{Pedrosa2015}
{Pedrosa}, S. \& {Tissera}, P. 2015, ArXiv e-prints

\bibitem[{{Pedrosa} {et~al.}(2010){Pedrosa}, {Tissera}, \&
  {Scannapieco}}]{Pedrosa2010}
{Pedrosa}, S., {Tissera}, P.~B., \& {Scannapieco}, C. 2010, \mnras, 402, 776

\bibitem[{{Pedrosa} {et~al.}(2014){Pedrosa}, {Tissera}, \& {De
  Rossi}}]{Pedrosa2014}
{Pedrosa}, S.~E., {Tissera}, P.~B., \& {De Rossi}, M.~E. 2014, \aap, 567, A47

\bibitem[{{Peebles}(1969)}]{peebles1969}
{Peebles}, P.~J.~E. 1969, ApJ, 155, 393

\bibitem[{{Romanowsky} \& {Fall}(2012)}]{romanowsky2012}
{Romanowsky}, A.~J. \& {Fall}, S.~M. 2012, ApJS, 203, 17

\bibitem[{{Sales} {et~al.}(2010){Sales}, {Navarro}, {Schaye}, {Dalla Vecchia},
  {Springel}, \& {Booth}}]{sales2010}
{Sales}, L.~V., {Navarro}, J.~F., {Schaye}, J., {et~al.} 2010, \mnras, 409,
  1541

\bibitem[{{Sales} {et~al.}(2012){Sales}, {Navarro}, {Theuns}, {Schaye},
  {White}, {Frenk}, {Crain}, \& {Dalla Vecchia}}]{Sales2012}
{Sales}, L.~V., {Navarro}, J.~F., {Theuns}, T., {et~al.} 2012, \mnras, 423,
  1544

\bibitem[{{Scannapieco} {et~al.}(2005){Scannapieco}, {Tissera}, {White}, \&
  {Springel}}]{scan2005}
{Scannapieco}, C., {Tissera}, P.~B., {White}, S.~D.~M., \& {Springel}, V. 2005,
  \mnras, 364, 552

\bibitem[{{Scannapieco} {et~al.}(2006){Scannapieco}, {Tissera}, {White}, \&
  {Springel}}]{scan2006}
{Scannapieco}, C., {Tissera}, P.~B., {White}, S.~D.~M., \& {Springel}, V. 2006,
  \mnras, 371, 1125

\bibitem[{{Scannapieco} {et~al.}(2009){Scannapieco}, {White}, {Springel}, \&
  {Tissera}}]{Scan2009}
{Scannapieco}, C., {White}, S.~D.~M., {Springel}, V., \& {Tissera}, P.~B. 2009,
  \mnras, 396, 696

\bibitem[{{Springel}(2005)}]{springel2005}
{Springel}, V. 2005, \mnras, 364, 1105

\bibitem[{{Springel} \& {Hernquist}(2003)}]{springel2003}
{Springel}, V. \& {Hernquist}, L. 2003, \mnras, 339, 289

\bibitem[{{Springel} {et~al.}(2001){Springel}, {Yoshida}, \&
  {White}}]{springel2001}
{Springel}, V., {Yoshida}, N., \& {White}, S.~D.~M. 2001, \na, 6, 79

\bibitem[{{Tissera} {et~al.}(2010){Tissera}, {White}, {Pedrosa}, \&
  {Scannapieco}}]{tissera2010}
{Tissera}, P.~B., {White}, S.~D.~M., {Pedrosa}, S., \& {Scannapieco}, C. 2010,
  \mnras, 406, 922

\bibitem[{{Tissera} {et~al.}(2012){Tissera}, {White}, \&
  {Scannapieco}}]{tissera2012}
{Tissera}, P.~B., {White}, S.~D.~M., \& {Scannapieco}, C. 2012, \mnras, 420,
  255

\bibitem[{{Woosley} \& {Weaver}(1995)}]{WW95}
{Woosley}, S.~E. \& {Weaver}, T.~A. 1995, \apjs, 101, 181

\end{thebibliography}

\end{document}